\documentclass[prl,twocolumn,showpacs,superscriptaddress,amsmath,amssymb,nofootinbib]{revtex4}
\usepackage{graphicx}
\usepackage{dcolumn}
\usepackage{bm}
\usepackage{slashed}
\usepackage{ulem} 
\usepackage{color}
\begin{document}
\title{Many-Body Atomic Speed Sensor in Lattices}

\author{Salvatore Marco Giampaolo}
\affiliation{International Institute of Physics, 59078-400 Natal-RN, Brazil}

\author{Andrea Trombettoni}
\affiliation{CNR-IOM DEMOCRITOS Simulation Center, Via Bonomea 265, I-34136 Trieste, Italy}
\affiliation{SISSA and INFN, Sezione di Trieste,
via Bonomea 265, I-34136 Trieste, Italy}

\author{Peter Kr\"uger}
\affiliation{Department of Physics and Astronomy, University of Sussex, 
Falmer, Brighton BN1 9RH, United Kingdom}

\author{Tommaso Macr\`i}
\affiliation{Departamento de F\'isica Te\'orica e Experimental,
Universidade Federal do Rio Grande do Norte, 59072-970 Natal-RN,Brazil}
\affiliation{International Institute of Physics, 59078-400 Natal-RN, Brazil}

\begin{abstract}
We study the properties of transmissivity of a beam of atoms traversing an optical lattice loaded with ultracold atoms.   
The transmission properties as function of the energy of the incident particles are dependent 
on the quantum phase of the atoms in the lattice.
In fact, in contrast to an insulator regime, the absence of an energetic gap in the spectrum of the superfluid 
phase enables the atoms in the optical lattice to adapt to the presence of the beam.
This induces a backaction process that has a strong impact on the transmittivity of the atoms. 
Based on the corresponding strong dependency we propose the implementation of a speed sensor with an estimated sensitivity of 
$10^8 - 10^9$m/s/$\sqrt{\rm Hz}$. 
We point out that the velocity sensitivity improves when the interaction term in the optical lattice increases.
Applications of the presented scheme are discussed.
\end{abstract}

\maketitle


{\it Introduction.}
Recent progress in the manipulation of atomic, molecular and optics systems in general, and quantum gases in particular~\cite{Bloch2008,Lewenstein2012}
forms the basis of a new class of quantum devices. 
A major line of research in this context is quantum sensing~\cite{Degen2017}, devoted to measurements enhanced or made possible by the low temperature,
low decoherence, and/or strong quantum correlations achieved in cold atom systems. 
Apart from the long-standing interest in the pursuit of increasing the performance of atomic clocks~\cite{Bloom2014}, 
one can foresee or perform measurements of accelerations and rotations~\cite{Muntinga2013,Stevenson2015}, 
and of other quantities -- see, {\it e.g.,} the recent proposal for the measurement of magnetic fields~\cite{Jachymski2017}.
Based on the application of a variety of interferometric schemes~\cite{Cronin2009,Tino2014}, the main advantages are that such devices may perform 
better at the micrometer scale, and that -- even though not the best in absolute sensitivity -- they can be portable~\cite{Fortagh2007}, with a 
breadth of research and technological applications. 

In this paper we illustrate a possible application of a new type of cold atom-based sensor that implements a speedometer. 
This sensor utilizes an atomic beam with low velocity spread colliding with an optical lattice loaded with ultracold bosonic atoms.
The interaction between the atoms in the beam and the ones in the lattice push latter atoms aside. 
In such a system therefore we can face a physical phenomenon that 
can refer as a sort of ``ultracold Moses effect''. 
This backaction creates resonances for the transmission of beam particles and 
makes the transmittivity of the beam of test particles extremely sensible even to small changes in the relative speed between the 
source of the beam and the lattice.
Therefore it can be used to realize a speed sensor in which, unlike most of other atom sensors, the interactions in the lattice 
may help to have larger sensitivities.

An essential ingredient of our proposal is the possibility to control a two-component cold gas. 
This has become a standard capability in quantum gas laboratories, where the two components can either be two hyperfine levels of a single species, 
or two different species. 
Applications of bosonic two-component gases range from the study of component separation in binary mixtures~\cite{Hall1998} and of the motion of 
impurities in Bose-Einstein condensates~\cite{Thalhammer2008} to Josephson tunneling induced by Rabi 
coupling~\cite{Williams2000,Smerzi2003,Cappellaro17}, high-resolution magnetometry~\cite{Vengalattore2007} and sub-shot-noise 
interferometry~\cite{Gross2010,Riedel2010}. 
Among the many experimental manipulation techniques it is possible to have different optical lattices acting on the two 
components~\cite{Mandel2003} and to confine them in different dimensionalities~\cite{Lamporesi2010}. 
In our case one of the two components is trapped in a lattice potential, while the other is propagating in a potential-free environment where the 
atoms are used as test particles. 
The colliding beam of atoms
is then directed to traverse the optical lattice. 
This is within experimental reach, as shown e.g.\ in~\cite{Gadway2012}, where a one-dimensional Bose gas was used as a source of matter waves to 
determine the spatial ordering of atoms of a different species confined in an optical lattice.  

\begin{figure}[t!]
\begin{center}
\includegraphics[width=0.45\textwidth,angle=0]{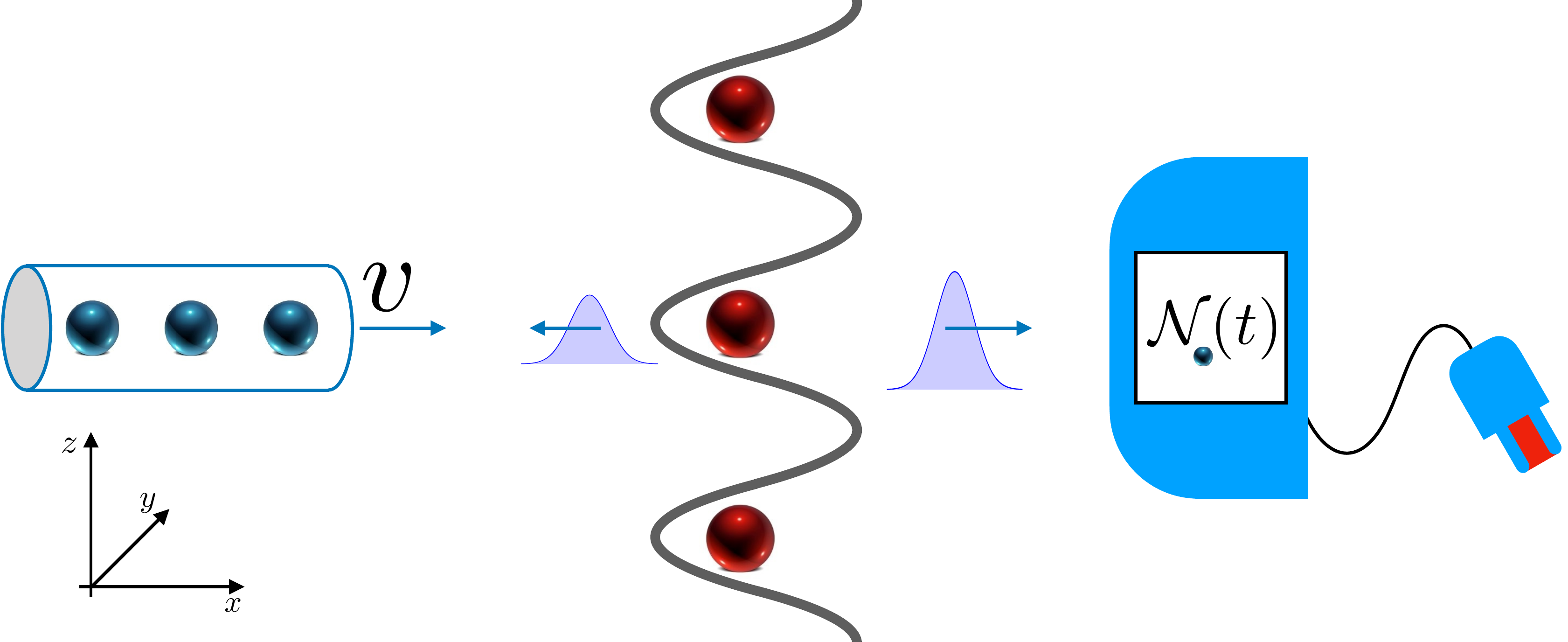}
\end{center}
\caption{Scheme of the speed sensor. A focused beam of test atoms (blue spheres) impacts a $1-$d
optical lattice in superfluid regime (red spheres).
The transmission rate of the colliding atoms depends on their speed respect to the lattice.
}
\label{fig0}
\end{figure}

{\it General considerations.}
The transmission rate $T(v)$ of a quantum particle across a region characterized by the presence of a (static) potential barrier depends on the 
kinetic energy of the object itself and thus, in the semi-classical regime, on its speed with respect to the barrier $v$.
However, the sensitivity of a measurement based on this dependency may be very low.
It is, however, possible to substantially increase the sensitivity of this approach by introducing a 
backaction, i.e. a feedback making the potential barrier able to adjust to a change of the kinetic energy of the test atoms. 
In our scheme, depicted in Fig.~\ref{fig0}, the role of the test particle is played by a focused one-dimensional beam of non-interacting atoms 
while the potential barrier is realized using an optical lattice loaded with ultracold bosons. 
We assume that 
{\it i)} the atoms in the lattice are in superfluid regime; 
{\it ii)} the beam of test atoms is centered on an individual site of the optical lattice and that, in the directions orthogonal to the propagation,
its profile is Gaussian;
{\it iii)} all interactions are local. 
Increasing (decreasing) the kinetic energy of the test atoms will result in a change of their penetration within the optical lattice 
(note that if the barrier has a peculiar form this may not be the case~\cite{Paris2017}).
This increased (decreased) penetration implies, as a consequence, the rising of site dependent potential in the optical lattice that induces a 
displacement of the atoms from the sites in which this potential is larger towards the ones in which is smaller.
The depletion in the sites impacted directly by the beam induces an effective reduction (enhancement) in the height of the potential barrier for 
the test atoms and therefore a faster increment (drop) of the transmittivity than in the case in which the migration is forbidden as, for example, 
in the Mott-insulator phase. 
As a consequence, in the superfluid regime the sensitivity is expected to be substantially increased.
In such scheme the role of the optical lattice is to concentrate the atoms with which the beam of test atoms interact leading to a relevant increment 
of the backaction phenomenon.

{\it Theoretical model.}
The Hamiltonian of bosonic atoms of mass $m_b$ in a one-dimensional optical lattice reads~\cite{Jaksch1998,Greiner2001,Bloch2005}
\begin{eqnarray}
 \hat H_{opt}&=& \int \mathrm{d}^3\mathbf{r}\hat \psi^\dagger(\mathbf{r}) 
\left[ -\frac{\hbar^2}{2 m_b} \nabla^2 + V_o(\mathbf{r}) \right] \hat \psi(\mathbf{r}) \nonumber \\
& & + \frac{g_0}{2} \int \mathrm{d}^3\mathbf{r}\,\hat \psi^\dagger(\mathbf{r})\hat \psi^\dagger(\mathbf{r})
\hat \psi(\mathbf{r})\hat \psi(\mathbf{r}),
\end{eqnarray}
where $V_o({\bf r})\!\!\!\!=\!\!\!\!V_0 \sin(\pi z/a)^2$ is the periodic potential and $g_0\!\!=\!\!4 \pi \hbar^2 a_{bb}/m_b$ the interaction 
strength of atoms in the lattice.
When the filling is small one can use a harmonic approximation for the Wannier wave-functions~\cite{Pilati2012,footnote1}, setting 
\mbox{$\hat \psi(\mathbf{r})=\sum_i \eta_i(\mathbf{r}) \hat b_i = \sum_i \eta_x(x) \eta_y(y) \eta_z(z-z_i)\, \hat b_i$},
where $\hat b_i$ is the bosonic annihilation operator on the \mbox{$i$-th} site in $(0,0,z_i)$ and 
$\eta_\alpha (\alpha)= \frac{1}{\pi^{1/4} \ell_\alpha^{1/2}} \exp\left(-\frac{\alpha^2}{2 \ell_\alpha^2}\right)$.
The  harmonic oscillator length $\ell_\alpha$ depends on the direction $\alpha$. 
In the lattice direction \mbox{$\ell_z=a/(\pi^4 V_o/E_r)^{1/4}$}, where \mbox{$E_r=\frac{\pi^2 \hbar^2}{2 m_b a^2}$} is the recoil energy 
and $a$ is the lattice spacing. In the orthogonal directions, \mbox{$\ell_{x,y}=\sqrt{\hbar/(m_b\, \omega_\perp)}$} 
depends on the frequencies of the harmonic trap that we assume to be equal  $\omega_x\equiv\omega_y=\omega_\perp$.
In a single-band approximation for the atoms in the lattice we recover the standard Bose-Hubbard model in the tight-binding limit that in the 
grand canonical ensemble becomes
\begin{equation}
\label{free_hamiltonian}
\hat H_\text{opt} \!= \!-t\!\sum_i (\hat b_i^\dagger \hat b_{i+1}+h.c.) + \frac{u}{2}\!\sum_i \hat n_i(\hat n_i-1)\!-\!\mu\! \sum_i \hat n_i,
\end{equation}
where $\hat{n}_i=\hat b_i^\dagger \hat b_i$, $u$, $\mu$ and $t$ are the intensity of 
the local intra-species interaction, the chemical potential and the hopping term, respectively.

The interaction of lattice bosons with beam particles is described by $\hat{H}_\text{int}$:
\begin{equation}
\label{definitionWi}
\hat{H}_{\text{int}} = g_{bt} \int \mathrm{d}^3\mathbf{r}\,\hat \phi^\dagger(\mathbf{r})\hat \psi^\dagger(\mathbf{r})
\hat \psi(\mathbf{r})\hat \phi(\mathbf{r}),
\end{equation}
where $g_{bt}= 2 \pi \hbar^2 a_{bt}/\mu_{bt}$ is the inter-species contact interaction and \mbox{$\mu_{bt}=m_b m_t/(m_b+m_t)$} the 
reduced mass of a test and a bosonic atom in the lattice.
We assume that $\hat \phi(\mathbf{r})=\chi(\mathbf{r}) \hat c=\chi_x(x) \chi_y(y) \chi_z(z)\, \hat c$ can be factorized in the three directions,
where $\hat c$ is the annihilation operator acting of the beam atoms. 
Test particles propagate along the $x$ direction and $\chi_y(y)$ and $\chi_z(z)$ are Gaussians with oscillator length $\ell_t$ fixed a priori 
and centered, respectively, around $y=0$ and $z=z_{i_0}$. 
$\chi_x(x)$ is the solution of the time independent one dimensional Schr\"odinger equation in the presence of a potential barrier
generated by the optical lattice
\begin{eqnarray}
 \label{Potential_barrier}
\!\! W(x) &=& \frac{2 \pi \hbar^2 a_{bt}}{\mu_{bt}} |\eta_x(x)|^2  \int dy |\chi_y(y)|^2 |\eta_y(y)|^2 \times \\
 \!\! & & \!\times \! \sum_i \langle \hat{n}_i \rangle \int dz |\chi_z(z-z_{i_0})|^2 |\eta_z(z-z_i)|^2 \nonumber
\end{eqnarray}
The full Hamiltonian of the atoms in the lattice then becomes
\begin{equation}
\label{interaction_hamiltonian}
\hat H=\hat H_\text{opt} + \sum_i \overline{W}_i \hat n_i\, \langle \hat{n}_{test} \rangle
\end{equation}
where $\hat{n}_{test}=c^\dagger c$ counts the particles number of the beam that interact simultaneously with the optical
lattice and \mbox{$\overline{W}_i=g_{bt} \int \mathrm{d}^3\mathbf{r}\, |\chi(\mathbf{r})|^2 |\eta_i(\mathbf{r})|^2 $}. 
In the stationary condition, i.e. when the flux of the particles in the beam as well as the lattice density distribution are constant, we can 
replace $\hat{n}_{test}$ with its average value $\langle \hat{n}_{test} \rangle$ within a mean field approximation.

{\it Numerical solution.} 
To determine the stationary state given by the solution of our problem we use the site-dependent mean field approach described in 
Refs.~\cite{Buonsante2008,Buonsante2009}.
The advantage of this approach is to vary the value of $\mu$ self consistently keeping the number of atoms in the lattice fixed and avoiding 
the depletion effect of standard mean field approaches.
As all the others mean field approaches it does not take into account quantum fluctuations and to reduce this problem we consider 
parameters far from the quantum critical point.
We implement such a method in five steps.
\textit{Step 1}: We find the mean field order parameter $\langle \hat b_i \rangle$ assuming $\overline{W}_i=0 \; \forall \;i$.
\textit{Step 2}: We determine $W(x)$ of eq.~(\ref{Potential_barrier}) and numerically solve the Schr\"odinger equation for the atoms in the beam. 
\textit{Step 3}: We use the $\chi_x(x)$ obtained in \textit{Step 2} to determine $\overline{W}_i$ and hence the new set of order parameters. 
\textit{Step 4}: We determine the total number of atoms in the optical lattice adjusting the chemical potential.
\textit{Step 5}: Finally we iterate \textit{Steps 2-4} until all quantities, i.e., the set of site dependent mean field order 
parameters, the total number of atoms in the lattice and the transmittivity converge (up to $10^{-8}$). 
\begin{figure}[t!]
\begin{center}
\includegraphics[width=0.47\textwidth,angle=0]{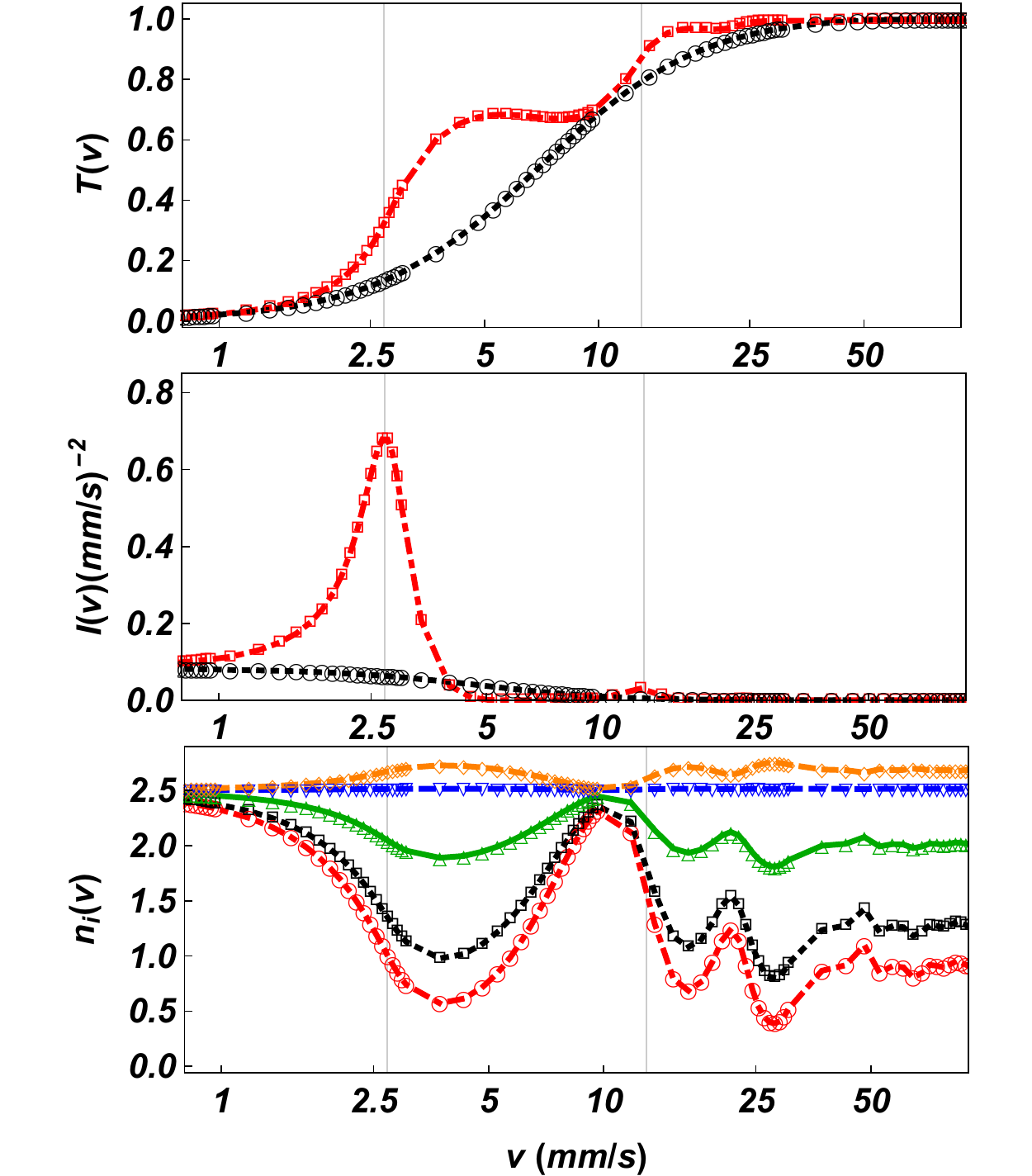}
\end{center}
\caption{
Behavior of $T(v)$ ({\it top}) and FI $I(v)$ ({\it central}).
Black dashed line (empty black circles): Mott-insulator like phase. Red dash-dotted solid line (red empty squares): superfluid phase.
{\it (Bottom)}. Average occupancy $n_i(v)=\langle \hat{n}_i \rangle$ of the optical lattice sites in the superfluid 
regime for several different sites:
red circle/dashed line $i=i_0$; black square/dotted line $i=i_0 \pm 1$;
green triangle up/solid line $i=i_0 \pm 2$; blue triangle down/dot dashed line $i=i_0 \pm 3$; 
orange diamond/dashed line $i=i_0 \pm 4$.
All quantities are plotted as function of the velocity of the test atoms in the reference system in which the optical lattice is at rest. 
Simulations were performed fixing $\omega_\perp=5$ kHz, $a=266$ nm and 
$V_0=7\,E_r$. The lattice is loaded with $^{87}$Rb with an average $2.5$ atoms per site and $a_{bb}=100$a$_0$.
Beam atoms are $^7$Li with $r\!=\!\ell_t/\ell_z\!=\!10$ and $a_{bt}=200$ a$_0$.
For these values $v_R=8.63$ mm/s. Gray vertical lines: position of local peeks of the FI.}
\label{fig1}
\end{figure}

{\it Fisher Information (FI) and sensitivity.}
From the knowledge of the transmission coefficient $T(v)$ we determine the optimal sensitivity via the FI 
$I(v)$~\cite{Frieden2004,Refregie2012r,Wasak2016}.
Experimentally, we have access to the flux of incoming particles and to the fraction of detected particles. 
Therefore we can define the probability of a single particle with velocity $v$ 
to be transmitted across the optical lattice as 
the transmissivity $P(1;v) \equiv T(v)$  while the probability to be reflected equals $P(0;v) \equiv 1-T(v)$. 
The resulting FI then reads
\begin{equation}
 \label{fisher_information}
 I(v)=\left(\frac{\partial T(v)}{\partial v}\right)^2 \frac{1}{T(v)(1-T(v))}.
\end{equation}
FI is strictly connected with the relative sensitivity $\sigma(v)$ that is the ratio between the 
relative change in the output signal  
and the relative change in the input 

\begin{equation}
 \sigma(v)=\frac{\Delta T(v)}{T(v)} \cdot \left(\frac{\Delta v}{v}\right)^{-1}\simeq v\sqrt{\frac{1-T(v)}{T(v)}I(v)} .
\end{equation}
Setting our system in such a way that, when the source of the beam is at rest with respect to the optical lattice, the velocity of test atoms
equals $v_m$, i.e. the velocity at which $I(v)$ reaches its maximum, we may determine velocities of the order of 
\begin{equation}
 \label{detectable_velocity}
v'=\frac{\Delta T}{T(v_m)} \sqrt{\frac{1}{ I(v_m)}\frac{T(v_m)}{1-T(v_m)}}.
\end{equation}

{\it Results.}
In the upper panel of Fig.~\ref{fig1} we show a typical example of $T(v)$ of a beam of test atoms across an optical lattice in the 
superfluid regime compared with the case in which the 
backaction effect is artificially suppressed~\cite{note}. 
For very slow and very fast particles, the two transmittivity coincide.
When the kinetic energy of test atoms, in a reference system in which the optical lattice is at rest, is comparable with the height of the potential 
barrier, the penetration of such atoms in the lattice is relevant. 
This affects the Hamiltonian seen by the atoms in the lattice and induces a redistribution of the atoms in the lattice that strongly depends on the 
velocity.
As a consequence we have a very fast increment of the transmission rate even with a small variation of $v$.
However such trend is not monotonic. 
Once the wave function has penetrated throughout the optical lattice, the spatial dependence of $\chi_x(x)$ can reduce the 
value of the integral in eq.~(\ref{definitionWi}).
This may compensate the natural increment of the transmission rate associated to an increment of the speed of the test atoms, hence generating the 
plateau in the transmission rate of the superfluid case (the upper panel of Fig.~\ref{fig1}). 
Increasing $g_{bt}$, the plateau can be replaced by a sharp downhill of $T(v)$.
In the presence of such plateau, as the FI depends on the square of the derivative of $T(v)$ respect to $v$, the sensitivity becomes smaller 
than the one in the Mott-Insulator regime.
The non monotonic behavior of the integral in eq.~(\ref{definitionWi}) induces also the modulation of the occupancy of the sites
(the lowest panel of Fig.~\ref{fig1}.)
The number of sites affected by the depletion depends on $r=\ell_t/\ell_z$. 
If $r \simeq 1$, then the depleting effect is almost contained in one single site, i.e. $i_0$, whereas the number of sites affected by the depletion 
increases when $r$ gets larger.

\begin{figure}[t!]
\begin{center}
\includegraphics[width=0.47\textwidth,angle=0]{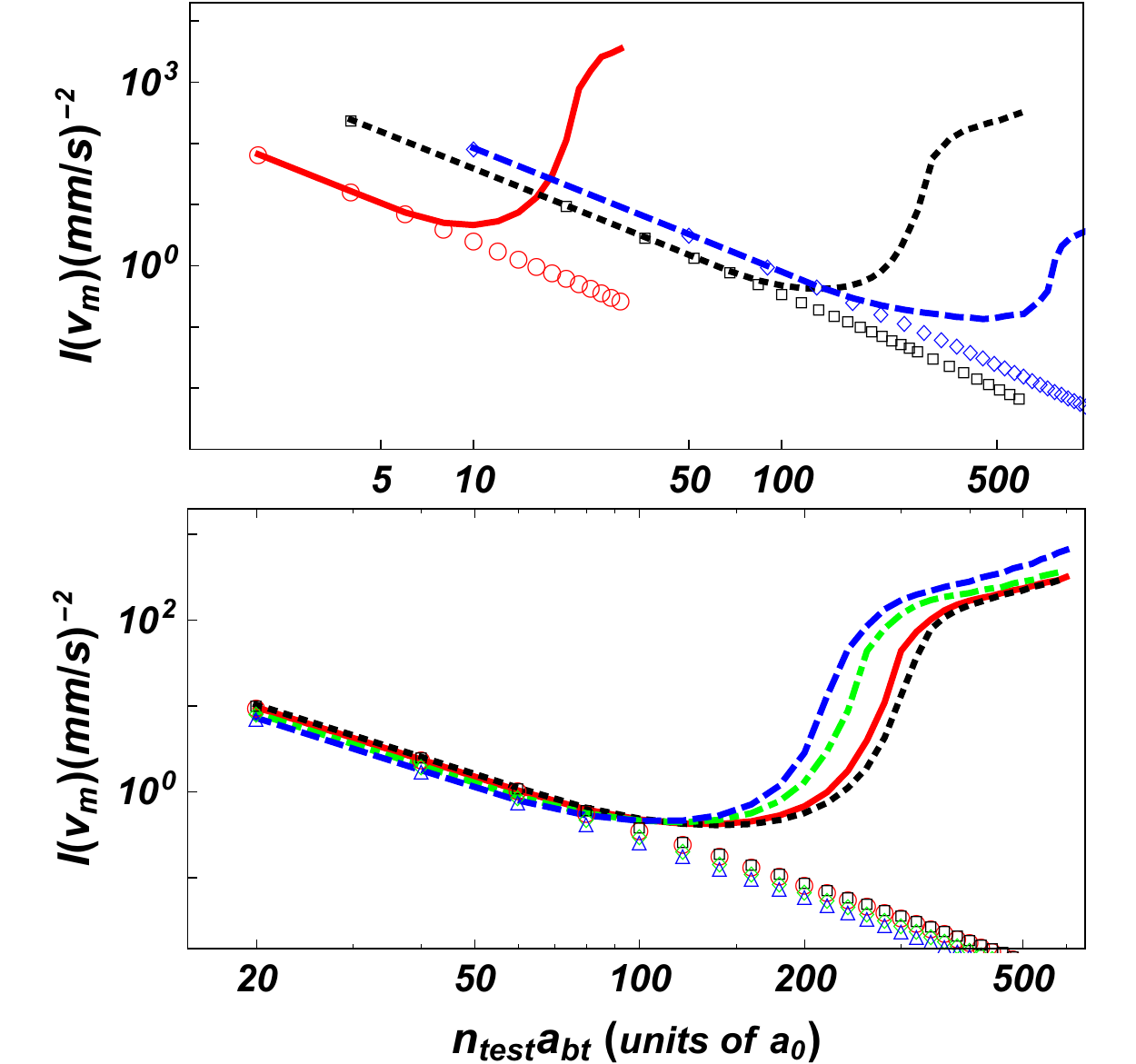}
\end{center}
\caption{
Maximum of FI as a function of the $\langle \hat{n}_{test} \rangle a_{bt}$ for different values of $r=\ell_t/\ell_z$ ({\it top})
and of the lattice depth $V_0$ ({\it bottom}). 
Lines: superfluid phase. Dots: Mott-Insulator like phase. 
{\it Top.} Lattice depth $V_0=7\,E_r\; (t/u \simeq 0.29)$ for: $r=2$ 
(red solid line/red empty circles); $r=10$ (black dotted line/black empty squares); $r=15$ (blue dashed line/blue empty diamonds).
{\it Bottom}. We fix $r=10$ and vary $V_0$: $V_0=6\, E_r$ ($u=0.048 E_r$, $t/u \simeq 0.42$) 
(black dotted line/black empty squares); $V_0=7\,E_r$ ($ u=0.050 E_r$, $ t/u \simeq 0.29$) (red solid line/red empty circles); 
$V_0=10\, E_r$ ($u=0.054 E_r$, $ t/u \simeq 0.11$) (green dot-dashed line/green empty diamonds); 
$V_0=13\, E_r $ ($u=0.058 E_r $, $t/u \simeq 0.04$) (blue dashed line/blue empty triangles).
Other parameters as in Fig.~\ref{fig1}.}
\label{fig2}
\end{figure}

The height and the position of the maximum of the FI, $I(v_m)=\max(I(v))$, depend on the parameters of the system. 
In Fig.~\ref{fig2} we show the dependence of $I(v_m)$ on the mean-field inter-species interaction $\langle \hat{n}_{test} \rangle a_{bt}$, 
for several value of $r=\ell_t/\ell_z$ (upper panel) and of the intensity of the optical lattice $V_0$. 
In all the cases analyzed, when the interaction between the two species of atoms is weak, the value of $I(v_m)$ coincides with the one obtained in the 
Mott-Insulator like limit. 
However, increasing $\langle \hat{n}_{test} \rangle a_{bt}$ above a certain threshold value, we have a pronounced increment of $I(v_m)$.  
This increment can be enhanced both reducing the width of the test beam (upper panel of Fig.~\ref{fig2}) and/or increasing the depth of the optical 
lattice, that implies a reduction of the ratio $t/u$ taking care to avoid to enter in the Mott-insulator regime.
If one has an integer filling and enter in the Mott Insulator phase one finds that the sensitivity decreases always linearly as in the limit
of low energies in Fig.~\ref{fig2}.

Summarizing, it is possible to obtain a value of $I(v_m)$ of the order of $10^{3}$--$10^4\;\mathrm{(mm/s)}^{-2}$ where $v_m$ is of the 
order of $1-10$ mm/s. 
The sensitivity of an atom detector can be estimated in the shot-noise limit by the fact that the error of the number of particles is 
$\propto \sqrt{N}$: therefore $\frac{\Delta T}{T} \propto \frac{1}{\sqrt{N}}$. 
Assuming an integration time of $1$ s and $\frac{\Delta T}{T} \sim 10^{-3}$, we obtain from eq.~(\ref{detectable_velocity}) that our system can 
measure velocities of the order $10^{-8}-10^{-9}$ m/s.
Generally with an integration time $\tau_{int}$ and a constant incoming flux $\dot N$, then $N=\dot N \, \tau_{int}$.
With $\dot N =10^6$/s we expect a sensitivity of $10^{-8}-10^{-9}$ m/s/$\sqrt{\text{Hz}}$. 
The distance over which the velocity is measured can be large (even $\sim$ cm) as long as the beam remains focused to within no more than a 
few lattice sites. 

We observe that using the technology in LIGO one can detect much smaller velocities, {\it e.g.} of order of $10^{-16}$ m/s. 
In standard atom interferometers one has $\sim 10^{-13}g$ as acceleration sensitivity, which in portable devices is $\sim 10^{-9}g$. 
For the latter one can therefore measure velocities of order of $10^{-9}$ m/s. 
Notice, however that in atom gravimeters to measure the velocity one typically has to have or assume 
constant acceleration, while in our scheme there is not such assumption.


{\it Further considerations.} 
A first issue to be discussed is the role of temperature on the proposed scheme. 
A simple estimate can be performed in self consistent harmonic approximation~\cite{Simanek1994}, showing that when the temperature $\tau$
is much smaller than the Bose-Einstein condensate critical temperature $T_{BEC}$ (e.g. $\tau \simeq 0.3\, T_{BEC}$) the 
effective value of $T$ is renormalized to $T_{eff}(\tau)=T e^{-D_{ij}}$, where $D_{ij}$ 
is the expectation value of $(\theta_{i}-\theta_j)^2$, with $\theta_i$ being the phase of the superfluid in the $i$-th well. 
This shows that there is quantitative, but not qualitative, effect of thermal fluctuations. 

Another issue is that it is not yet possible to have a continuous coherent beam at degenerate temperature. 
Two remedies are presently possible: First, the beam can be replaced by a cloud trapped in a moving (parabolic) guiding potential. 
Second, the beam is continuous but at a temperature above $T_{BEC}$. 
This is acceptable as long as the beam is dilute enough not to heat (within the time of the experiment) the atoms in the lattice above their critical 
temperature for superfluidity. 
Note that when the velocity $v$ is on the order of $10^{-3}$ m/s and the number of particles in the beam 
is $M \sim 10^6$, then $M \frac{m}{2} v^2 \sim 10^{-3} nK$, i.e.\ much lower than typical temperatures in the relevant experiments. 
Finally, if there are fluctuations varying the intensity of the optical lattice potential 
$V_0$, one expects a variation of $t/u$, which it appears to not have a major effect on the speed sensitivity, as shown in Fig.~\ref{fig2}. 
Three-body losses may vary the total number of particles in the lattice, reducing the time 
on which one can perform a measurement and the minimum detectable velocity, but not altering significantly the sensitivity.
Further work will be devoted to a deeper analysis of detrimental effects for a realistic implementation of the proposal device.  

To conclude, our scheme can be useful to design a gyroscope. 
We can consider a ring shape optical lattice with a radius $R$ and a beam of test atoms propagating 
in the ring plane. The beam of test particle is tangential to the ring.
As above, the beam interacts relevantly only with a small number of sites of the lattice. 
In this realization a measurement of $v$ becomes a measurement of the rotational speed $v/R$. 
Considering the value that we have obtained so far we have that, for a radius $R \sim 10 \, \mu$m, 
one can then measure variations in the 
angular velocity of the order of $10^{-5}$ rad/s. 
Even if we expect that this implementation would not show a qualitative difference to what we presented here,
it would require a separate study to determine the sensitivity of rotation measurement into this geometry that is different to the one considered in the text.
Even though this application is certainly challenging, 
we believe it illustrates a possible 
application of this scheme for other measurements at 
micrometer scale.

{\it Conclusions.} We presented a scheme to perform sensitive velocity measurements based on an atomic beam impacting on atoms confined in an optical 
lattice. The sensitivity depends on a many-body 
backaction mechanism
determined by the probed particles of the 
atoms in the optical lattice.

\begin{acknowledgments}
{\it Acknowledgments}. We acknowledge A. Smerzi and all
members of the Matterwave consortium for useful discussions.
T.M. acknowledges CNPq for support through Bolsa de produtividade
em Pesquisa n. 311079/2015-6. Support from the EU-FET Proactive Action 
(Grant 601180 “MatterWave”) is also acknowledged.
\end{acknowledgments}


\begin{thebibliography}{*}
\bibitem{footnote1}
If the filling or the interaction is large, (or the transverse confinement is weaker than the one considered in the text), 
one should determine the $\eta$'s and the Bose Hubbard parameters $t$ and $u$ from the solution of the Gross-Pitaevskii equation with the
nonlinear term~\cite{Smerzi2003-2}.
\end{thebibliography}
\end{document}